\documentclass[epjST]{svjour}
\usepackage{graphics}
\usepackage{hyperref}
\usepackage{amsmath}
\usepackage[dvips]{epsfig}
\newcommand{\beq}{\begin{equation}}
\newcommand{\eeq}{\end{equation}} 
\newcommand{\beqa}{\begin{eqnarray}}
\newcommand{\eeqa}{\end{eqnarray}}
\newcommand{\ba}{\begin{array}}
\newcommand{\ea}{\end{array}}

\begin{document}

\title{Shock waves in quasi one-dimensional Bose-Einstein condensate}

\author{Luca Salasnich}

\institute{Dipartimento di Fisica ``Galileo Galilei'' and CNISM, \\
Universit\`a di Padova, Via Marzolo 8, 35131 Padova, Italy \\
CNR-INO, via Nello Carrara 1, 50019 Sesto Fiorentino, Italy \\
e-mail: luca.salasnich@unipd.it}

\abstract{
We study analytically and numerically the generation of shock waves 
in a quasi one-dimensional Bose-Einstein condensate (BEC) made of 
dilute and ultracold alkali-metal atoms. For the BEC we use an equation 
of state based on a 1D nonpolynomial Schr\"odinger equation (1D NPSE), 
which takes into account density modulations in the transverse 
direction and generalizes the familiar 1D Gross-Pitaevskii equation (1D GPE). 
Comparing 1D NPSE with 1D GPE we find quantitative differences 
in the dynamics of shock waves regarding the velocity of propagation, 
the time of formation of the shock, and the wavelength of 
after-shock dispersive ripples.}

\maketitle

\section{Introduction}

In the past years the zero-temperature hydrodynamic equations 
of superfluids \cite{landau} have been successfully 
applied to investigate dilute and ultracold bosonic 
superfluids, i.e. Bose-Einstein condensates (BECs) 
made of confined alkali-metal 
atoms \cite{bec-review}. More recently it has been shown \cite{pitaevskii} 
that also fermionic superfluids in the BCS-BEC crossover can be 
modelled by hydrodynamic equations and their 
generalizations \cite{luca12}. 
\par 
The formation of shock waves induced by an
{\it ad hoc} external perturbation has been 
observed in BECs \cite{hau,cornell,hoefer,davis} 
and in the unitary Fermi gas \cite{thomas}. 
Motivated by these low temperature experiments 
on shock waves in atomic superfluids, 
some authors have theoretically analyzed the formation of shock waves 
in various configurations of bosonic 
\cite{zak,damski,gammal,perez,muga,noi} 
and fermionic superfluids \cite{noi2a,noi2b,bulgac}. 
In particular, shock waves of a strictly one-dimensional (1D) superfluid 
in the regime of quasi condensation have been studied in 
Refs. \cite{zak,damski} by using the 1D Gross-Pitaevskii equation (1D GPE). 
\par
Here we extend their analysis by studying a cylindrically confined BEC,  
whose transverse width $\sigma$ is not frozen but depends 
on the longitudinal axial density $\rho$ \cite{npse,gll}, and adopting 
the 1D nonpolynomial Schr\"odinger equation (1D NPSE) \cite{npse}. 
Indeed, 1D NPSE is a more realistic model for a 
comparison with experiments, where 
quasi one-dimensional BECs are produced by imposing a very strong 
harmonic confinement of characteristic length $a_{\bot}$ 
along two directions. In this paper we find that the shock waves 
obtained by using 1D GPE, based on the hypothesis of a strictly 
one-dimensional Bose-Einstein condensate, show quantitative 
differences with respect to the ones derived from 1D NPSE. 

\section{Hydrodynamic equations for a quasi 1D bosonic superfluid}

We consider a dilute bosonic superfluid at zero temperature 
that is freely propagating in the longitudinal 
axial direction $z$ and is confined in the transverse directions 
$(x,y)$ by a harmonic potential of frequency $\omega_{\bot}$ 
and characteristic length $a_{\bot}=(\hbar/(m\omega_{\bot}))^{1/2}$ 
with $m$ the mass of a bosonic particle. 
Under the adiabatic hypothesis that the transverse profile of the 
superfluid has a Gaussian shape with a size depending on the 
longitudinal density \cite{npse,gll}, one finds the following 
dimensionless hydrodynamic equations for the axial dynamics of the superfluid 
\beqa 
{\partial \rho \over \partial t} + 
{\partial \over \partial z} 
\left( \rho v \right) = 0 \; , 
\label{hy1}
\\
{\partial v \over \partial t} + 
{\partial\over \partial z} 
\left( {v^2 \over 2} + \mu(\rho) \right) = 0 \; , 
\label{hy2}
\eeqa
where $\rho(z,t)$ is the superfluid axial density at time $t$, 
$v(z,t)$ is the superfluid axial velocity at time $t$ and 
$\mu(\rho )$ is the bulk chemical potential. Here length is in units 
of $a_{\bot}$, time in units of $\omega_{\bot}^{-1}$, energy in units 
of $\hbar \omega_{\bot}$, and axial density in units of $a_{\bot}^{-1}$. 
Thus the axial dynamics of the confined superfluid depends crucially 
on the functional form of the bulk chemical potential, 
that is given by \cite{gll} 
\beq 
\mu = {3\over 2} \ \rho^2 \ L[{g\over \rho \sigma^2}] + {1\over \sigma^2} \; , 
\label{zora1}
\eeq
where 
\beq 
\sigma^4 = 1 + g \ \rho \ L'[{g\over \rho \sigma^2}] 
\label{zora2}
\eeq
is the transverse width, $g=2a_s/a_{\bot}$ is the interaction strength 
with $a_s$ the repulsive ($a_s>0$) 
s-wave scattering length of the inter-atomic 
potential and $a_{\bot}$ is the harmonic length of the 
transverse external potential \cite{gll}. Here L[x] 
is the Lieb-Liniger function \cite{ll}, which is defined as the solution 
of a Fredhholm equation and is such that 
\beq 
L[x] = \left\{ 
\begin{array}{lcr}
x-{4\over 3\pi} x^{3/2} & \mbox{for} & x\ll 1 \\
{\pi^2\over 3} ({x\over x+2})^2 & \mbox{for} & x\gg 1 
\end{array}
\right. \; . 
\eeq

As discussed in Ref. \cite{gll}, under the condition $g<1$, 
for $g\rho \ll g^2$ one gets the 1D Tonks-Girardeau 
regime \cite{tg} where $\sigma=1$ and 
\beq 
\mu = {1\over 2} \pi^2 \rho^2 + 1 \;  
\eeq 
with $1$ the adimensional transverse energy. Instead, 
for $g\rho \gg g^2$ one finds the (mean-field) BEC regime 
where $\sigma=(1+g\rho)^{1/4}$ and 
\beq
\mu = {g \rho \over \sqrt{1 + g \rho}}
+ {1\over 2}
\left( \sqrt{1+g\rho} + {1\over \sqrt{1+g\rho}}
\right) \; . 
\label{zora3}
\eeq
In this BEC regime one can distinguish two sub-regimes: the 1D 
quasi BEC regime for $g^2 \ll g \rho \ll 1$ where 
$\sigma \simeq 1$ and $\mu = g \rho + 1$, and the 3D BEC regime 
for $g\rho \gg 1$ where $\sigma=(g \rho)^{1/4}$ and 
$\mu=\sqrt{2 g \rho}$ \cite{gll}. 

Notice that a different expression with respect to Eq. (\ref{zora3}) 
has been heuristically proposed in Ref. \cite{spagnoli}. 
However, the 1D nonlinear Schr\"odinger equation of Ref. \cite{spagnoli}, 
which is an alternative model with respect to NPSE 
to describe transverse effects, gives a 1D equation of state 
very close to the NPES one, Eq. (\ref{zora3}), and both are 
in very good agreement with the 3D equation of state 
of the full 3D Gross-Pitaevskii equation \cite{npse,spagnoli}. 
In Eq. (\ref{zora3}) the term $g \rho/\sqrt{1 + g \rho}$ is the axial 
chemical potential while 
$(1/2)(\sqrt{1+g\rho} + 1/\sqrt{1+g\rho})$ 
is the transverse chemical potential, where 
$(1+g\rho)^{1/4}$ is the transverse width of the 
superfluid. 

Eqs. (\ref{hy1}) and (\ref{hy2}) 
with Eq. (\ref{zora3}) can be derived \cite{npse,kamchatnov}
from the 3D Gross-Pitaevskii equation \cite{bec-review} by using 
a variational wavefunction with a transverse 
Gaussian shape \cite{comesempre}. 
They are remarkably accurate in describing static and 
dynamical properties of Bose-Einstein condensates under transverse 
harmonic confinement. Usually in Eq. (\ref{hy2}) there is also a 
quantum pressure (QP) term, given by $1/(2\sqrt{\rho}) 
{\partial^2 \sqrt{\rho}/\partial z^2}$, and in this case 
the hydrodynamic equations are equivalent to a nonlinear 
Schr\"odinger equation, the so-called nonpolynomial Schrodinger 
equation (NPSE) \cite{npse}. NPSE has been used to successfully model 
cigar-shaped condensates by many experimental 
and theoretical groups (see for instance \cite{toscani,tedeschi}). 
It has been also used to study the role of a varying 
transverse width \cite{npse2} in the Josephson effect of 
a Bose-Einstein condensate with double-well axial 
confinement \cite{giovanni}. 

The QP term is negligible as long as 
the longitudinal width of the density 
profile $\rho(z)$  is larger than the healing length 
$\xi(\rho)$. A simple estimation based on the comparison between 
QP term and interaction term shows that 
$\xi(\rho)= (1+g\rho)^{1/4}/\sqrt{2g\rho}$. As previously 
discussed, Eq. (\ref{zora3}) cannot describe the very-low-density 
regime of Tonks-Girardeau ($g\rho \ll g^2$, with $g<1$), 
where the system behaves as a 1D gas of impenetrable Bosons 
\cite{gll,ll,tg}. 

\section{Sound velocity and shock waves} 

First we observe that the sound velocity 
$c_s(\rho)$ can be obtained from the Eq. (\ref{zora3}) 
by using the thermodynamics formula $c_s^2 = 
\rho (\partial \mu/\partial \rho )$ and is given by 
\beq 
c_s (\rho ) = \sqrt{g\rho (1+{3\over 4} g \rho ) \over (1+ g\rho)^{3/2} } 
\; ,  
\label{nice-cs} 
\eeq 
see also \cite{luca-cs}. 
The sound velocity $c_s(\rho)$ is the speed of propagation 
of a small (infinitesimal) perturbation of the initial condition 
of axially homogeneous system. By using $c_s(\rho)$ 
the two hydrodynamic equations (\ref{hy1}) and (\ref{hy2}) can be written 
in a more compact form as 
\beq 
{\partial \rho\over \partial t} + v {\partial \rho\over \partial x} 
+ {\partial v\over \partial x} \rho = 0 \; , 
\eeq
\beq 
{\partial v\over \partial t} + v {\partial v\over \partial x} + 
{c_s(\rho)^2 \over \rho} {\partial \rho\over \partial x} = 0 \; ,    
\eeq 
where dots denote time derivatives and 
primes space derivatives. 

As discussed by Landau and Lifshits \cite{landau}, 
exact solutions of these equations can be found 
by imposing that the velocity $v$ depends explicitly 
on the density $\rho$. In this way one has ${\partial v\over \partial t}=
{\partial v \over \partial \rho} {\partial \rho\over \partial t}$,  
${\partial v\over \partial x}
={\partial v \over \partial \rho} {\partial \rho\over \partial x}$ and 
the hydrodynamic equations become 
\beq 
{\partial \rho\over \partial t} + v {\partial \rho\over \partial x} 
+ {\partial v \over \partial \rho} 
{\partial \rho\over \partial x} \rho = 0 \; , 
\label{h1}
\eeq 
\beq 
{\partial v \over \partial \rho} {\partial \rho\over \partial t} 
+ v {\partial v \over \partial \rho} {\partial \rho\over \partial x} 
+ {c_s(\rho)^2 \over \rho} {\partial \rho\over \partial x} = 0 \; .  
\label{h2}
\eeq 
We now impose that the two equations reduce to 
the same hyperbolic equation 
\beq 
{\partial \rho\over \partial t} 
+ c(\rho) {\partial \rho\over \partial x} = 0 \; , 
\eeq 
where 
\beq 
c(\rho ) = v(\rho) + {\partial v \over \partial \rho}\rho 
\label{cc1}
\eeq
from Eq. (\ref{h1}), but also 
\beq 
c(\rho) = v(\rho) + {c_s(\rho)^2 \over \rho} 
\left({\partial v \over \partial \rho} \right)^{-1} \; ,  
\label{cc2}
\eeq 
from Eq. (\ref{h2}). 
It is quite easy to verify that, given a initial condition 
$f(z)$ for the density profile, the time-dependent solution $\rho(z,t)$ 
of the hyperbolic equation satisfies the following implicit equation: 
\beq 
\rho(z,t) = f (z - c(\rho(z,t)) t) \; ,  
\eeq 
which holds for regular solutions. 
In general, the initial wave packet $f(z)$ splits into two pieces 
travelling in opposite directions depending on the sign of $c(\rho)$. 
For simplicity in the following 
we consider the right-moving part only. 

To determine $c(\rho)$ we observe that, from the equality of Eqs. (\ref{cc1}) 
and (\ref{cc2}) we get 
\beq 
{\partial v \over \partial \rho}\rho = 
{c_s(\rho)^2 \over \rho} 
\left( {\partial v \over \partial \rho} \right)^{-1} \; , 
\eeq
from which 
\beq 
{\partial v \over \partial \rho} = {c_s(\rho) \over \rho}   \; , 
\label{rompino}
\eeq
and 
\beq 
v(\rho) = \int_{\rho_0}^{\rho}  {c_s({\tilde \rho}) \over {\tilde \rho} }  
d{\tilde \rho} \; ,  
\eeq 
where we impose that at infinity the initial 
density is constant, i.e. $\rho_0=f(z=\pm \infty)$, and  
the initial velocity field is zero, i.e. $v(\rho_0)=0$. 
Notice that in Eq. (\ref{rompino}) we have chosen 
${\partial v/\partial \rho}>0$ on physical grounds and 
that the relation between the velocity and density in Eq. (18)  
is precisely Eq. (101.4) of Ref. \cite{landau}. 
This formula can be integrated by using Eq. (\ref{nice-cs}). 
One finally has 
\beq 
v(\rho ) = \Phi(\rho ) - \Phi(\rho_0) \; , 
\label{nice-v}
\eeq 
where 
\beq 
\Phi(\rho )= 2 \sqrt{g \rho} 
F_1\left[{1\over 2},-{1\over 2},{3\over 4},
{3\over 2},-{3\over 4}g\rho,-g\rho\right] \; .  
\label{F1}
\eeq
The function $F_1[a,b_1,b_2,c,x,y]$ is the Appel 
hypergeometric function given by 
\beq 
F_1[a,b_1,b_2,c,x,y] = 
\sum_{m=0}^{\infty} \sum_{n=0}^{\infty}
{(a)_{m+n} (b_1)_m (b_2)_n \over (c)_{m+n} m! n!} x^m y^n \; , 
\eeq 
where $(a)_{m}=\Gamma(a+m)/\Gamma(a)$ is the Pochhammer 
symbol with $\Gamma(x)$ the Euler gamma function. 
The velocity $c(\rho)$ follows directly from the velocity 
$v(\rho)$ by using Eq. (11) and Eq. (14). It reads 
\beq 
c(\rho ) = v(\rho) + c_s(\rho) = 
\Phi(\rho ) - \Phi(\rho_0) + c_s(\rho ) \; . 
\label{nice-c}
\eeq 
Eqs. (\ref{nice-v}) and (\ref{nice-c}) with $\Phi(\rho)$ given by 
(\ref{F1}) are the main results of the paper. Note that 
in the 1D regime ($g\rho \ll 1$), where 
$c_s(\rho )=\sqrt{g\rho}$, one finds 
\beq 
v(\rho ) = 2 \sqrt{g \rho} - 2 \sqrt{g \rho_0} \; ,  
\label{v1d}
\eeq 
and 
\beq 
c(\rho ) = 3 \sqrt{g \rho} - 2 \sqrt{g \rho_0} \; , 
\label{c1d}
\eeq
which are the results obtained by Damski \cite{damski}. 

\begin{figure}
\begin{center}
{\includegraphics[width=8.cm,clip]{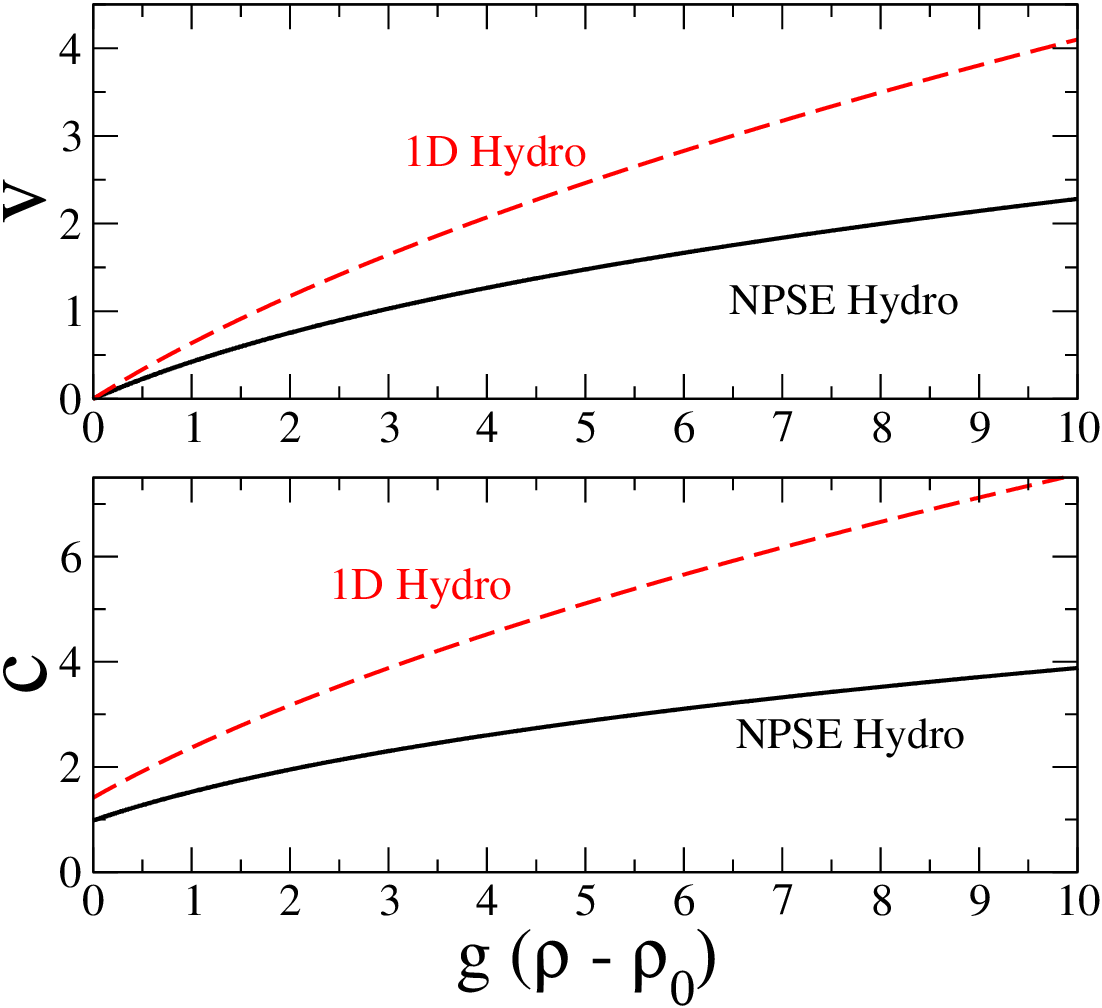}}
\caption{Velocities $v$ and $c$ as a function 
of the scaled axial density $g \rho$, with $g\rho_0=2$. 
Solid lines: Eqs. (\ref{nice-v}) and (\ref{nice-c}). 
Dashed lines: Eqs. (\ref{v1d}) and (\ref{c1d}).} 
\label{fig1}
\end{center}
\end{figure}

In Fig. \ref{fig1} we plot the velocities $v$ and $c$ as a function 
of the scaled and shifted axial density $g (\rho-\rho_0)$, chosing 
$g\rho_0=2$ to enhance the differences between 1D and NPSE hydrodynamics. 
The solid lines are obtained by using the NPSE hydrodynamics, 
i.e. Eqs. (\ref{nice-v}) and (\ref{nice-c}), while 
the dashed lines are obtained by using the 1D hydrodynamics, 
i.e. Eqs. (\ref{v1d}) and (\ref{c1d}). The two panels clearly show 
quantitative differerences between 1D and NPSE hydrodynamics. Notice that  
$v(\rho_0)=0$ while $c(\rho_0)=c_s(\rho_0)$.

\section{Formation of the shock wave front} 

Up to now the initial shape of the wave has been arbitrary. 
We consider now an example by choosing 
the following initial density profile 
\beq 
f(z) = \rho_0  + \rho_0 \; \eta \; e^{-z^2/(2\sigma)^2}  \; , 
\label{initial-rho}
\eeq
where $\eta$ describes the maximum impulse with respect 
to the density background $\rho_0$. 
Both amplitude $\mathcal{A}(\eta )$ and velocity $\mathcal{V}(\eta )$
of the impulse maximum are constant during time evolution. 
The amplitude is $\mathcal{A}(\eta )=\rho_0(1+\eta)$ while the 
velocity reads
\beq 
{\mathcal{V}}(\eta ) = c(\rho_0 (1+\eta)) \; . 
\eeq
As expected, taking $\eta =0$ the velocity of the impulse 
maximum reduces to the sound velocity $c(\rho_0)=c_s(\rho_0)$. 
Moreover, bright perturbations ($\eta >0$) move 
faster than dark ones ($\eta < 0$) \cite{noi2a}. 
Let us consider a bright perturbation. The speed of 
impulse maximum is bigger than the speed of its tails. 
As a result the impulse self-steepens in the direction 
of propagation so that the formation of a shock wave front 
takes place. The time $T_s$ required for such a process can 
be estimated as follows: the shock wave front appears 
when the distance difference traveled by lower and upper impulse 
parts is equal to the impulse half-width $2\sigma$, 
namely $ [{\mathcal{V}}(\eta ) - {\mathcal{V}}(0)] T_s = 2 \sigma $. 
It gives 
\beq 
T_s = {2 \sigma \over c(\rho_0 (1+\eta)) - c_s(\rho_0) } \; ,  
\label{nice-ts}
\eeq 
where the local velocity $c(\rho)$ is given by Eq. (\ref{nice-c}) 
while the sound velocity $c_s(\rho)$ is given by Eq. (\ref{nice-cs}). 
In the 1D regime ($g\rho \ll 1$) the formula of the time $T_s$ reads
\beq 
T_s = {2 \sigma \over 3 (\sqrt{1+\eta}-1)\sqrt{g\rho_0} } \; .    
\label{ts1d}
\eeq 
In the case of a dark perturbation ($\eta <0$) the tails  
of the wave packet move faster than the impulse minimum 
and the time of shock formation is simply 
$T_s = 2 \sigma/(c_s(\rho_0) - c(\rho_0(1+\eta))$. 

\begin{figure}
\begin{center}
{\includegraphics[width=8.cm,clip]{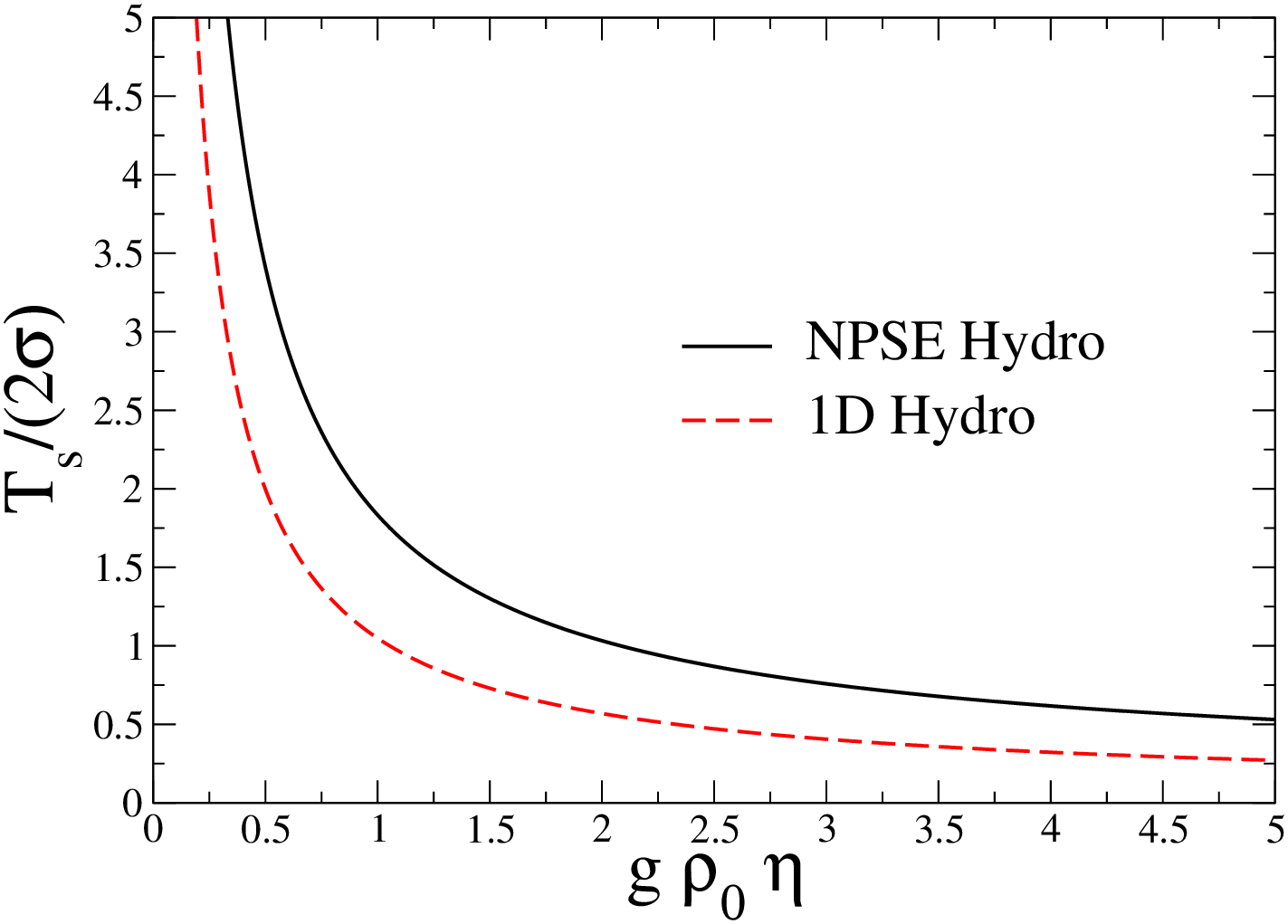}}
\caption{Scaled time of shock formation $T_s/(2\sigma)$ as function 
of the scaled impulse maximum $\rho_0 \eta$. 
$\sigma$ is the initial width of the impulse and we set $g\rho_0=2$. 
Solid line: Eq. (\ref{nice-ts}). 
Dashed line: Eqs. (\ref{ts1d}). } 
\label{fig2}
\end{center}
\end{figure}

In Fig. \ref{fig2} we plot the scaled time $T_s/(2\sigma)$ 
of shock formation as a function of the scaled impulse maximum $\rho_0 \eta$. 
Again we choose $g\rho_0=2$. 
The solid line is obtained by using the NPSE hydrodynamics, 
i.e. Eq. (\ref{nice-ts}), while 
the dashed line is obtained by using the 1D hydrodynamics, 
i.e. Eq. (\ref{ts1d}). Both curves give $T_s/(2\sigma)\to +\infty$ 
as $g \rho_0 \eta \to 0$. 
\par 
It is important to stress that our analytical results have been 
obtained without taking into account the quantum pressure 
term $1/(2\sqrt{\rho}) {\partial^2 \sqrt{\rho}/\partial z^2}$ 
in Eq. (2). Initially the spatial scale of density variations 
is $\sigma$, which can be chosen greater than the bulk healing length 
$\xi(\rho_0) = (1+g\rho_0)^{1/2}/\sqrt{2g\rho_0}$. As a result the QP term is 
negligible at the begining of time evolution. 

\section{Numerical results and after-shock dynamics} 

As the shock forms up density modulations occur on smaller and smaller 
length scales beeing finally of the order of the healing 
length. Damski \cite{damski} has numerically shown, for the strictly 
1D Bose gas, that the effect of quantum pressure 
term is that of preserving the single-valuedness 
of the density profile by inducing density oscillations at the 
shock wave front. Thus, after the formation of the shock 
Eqs. (\ref{hy1}) and (\ref{hy2}) are not reliable. 
To overcome this difficulty we include the dispersive quantum pressure term 
in the hydrodynamic equations, which become 
\beqa 
{\partial \rho \over \partial t} + 
{\partial \over \partial z} 
\left( \rho v \right) = 0 \; , 
\label{hy1g}
\\
{\partial v \over \partial t} + 
{\partial\over \partial z} 
\left( {v^2 \over 2} + \mu(\rho) 
- {1\over 2 \sqrt{\rho}} {\partial^2 \over \partial z^2} 
\sqrt{\rho} \right) = 0 \; .  
\label{hy2g}
\eeqa
We stress that at zero temperature for a viscousless superfluid 
the simplest regularization process of the shock is a purely 
dispersive quantum gradient term, 
which is proportional to $\hbar^2$ in dimensional units. Clearly, 
Eq. (\ref{hy2}) is a first order equation while Eq. (\ref{hy2g}) 
is not due to the quantum gradient term. 

Introducting the complex field $\psi(z,t)$ such that 
\beq 
\psi(z,t) = \sqrt{ \rho(z,t) } \, e^{i \theta(z,t)} \;  
\eeq
and 
\beq 
v(z,t) = {\partial\over \partial z} \theta(z,t) \; . 
\eeq
it is strightforward to show that Eqs. (\ref{hy1g}) and (\ref{hy2g}) are 
equivalent to the following one-dimensional nonlinear  
Schr\"odinger equation  
\beq 
i{\partial\psi \over \partial t} = \left[ -{1\over 2}{\partial^2\over 
\partial z^2} + \mu(|\psi|^2) \right] \psi \; . 
\eeq
Using the bulk chemical potential $\mu(\rho)$ of Eq. (\ref{zora1}) with 
(\ref{zora2}), this nonlinear Schr\"odinger equation is the time-dependent 
generalized Lieb-Liniger equation we introduced some years ago \cite{gll}
to accurately describe an experiment on a Tonks-Girardeau 
gas of $^{87}$Rb atoms \cite{weiss}. In the BEC regime, where 
Eq. (\ref{zora1}) reduces to Eq. (\ref{zora3}), the time-dependent 
generalized Lieb-liniger equation becomes the 1D time-dependent 
nonpolynomial Schr\"odinger equation (NPSE) 
\beqa 
i{\partial\psi \over \partial t}  &=& \Big[ -{1\over 2}{\partial^2\over 
\partial z^2} + {g |\psi|^2 \over \sqrt{1 + g |\psi|^2}}
\\
&+& {1\over 2} \left( \sqrt{1+g |\psi|^2} + {1\over \sqrt{1 + g |\psi|^2}}
\right)  \Big] \psi \; , 
\nonumber 
\label{tdnpse}
\eeqa
which gives the familiar one-dimensional Gross-Pitaevskii equation 
\beqa 
i{\partial\psi \over \partial t} &=& \Big[ -{1\over 2}{\partial^2\over 
\partial z^2} + g |\psi|^2 + 1 \Big] \psi \;  
\label{tdgpe}
\eeqa
in the 1D quasi BEC regime, where $g^2 \ll 
g|\psi|^2\ll 1$ and the transverse width $\sigma$ of Eq. (\ref{zora2}) 
becomes $\sigma \simeq 1$ (i.e. $\sigma \simeq a_{\bot}$ 
in dimensional units) \cite{npse}. 

We solve numerically Eq. (\ref{tdnpse}) by using a Crank-Nicolson 
finite-difference predictor-corrector algorithm \cite{sala-numerics} 
with the initial condition given by Eq. (\ref{initial-rho}) 
and $v(z,t=0)=0$. In fact, as also shown by Damski \cite{damski}, 
we have verified that the initial velocity field $v(\rho(z,t=0))$ and  
$v(z,t=0) = 0$ give practically the same time evolution. 

\begin{figure}
\begin{center}
{\includegraphics[width=7.cm,clip]{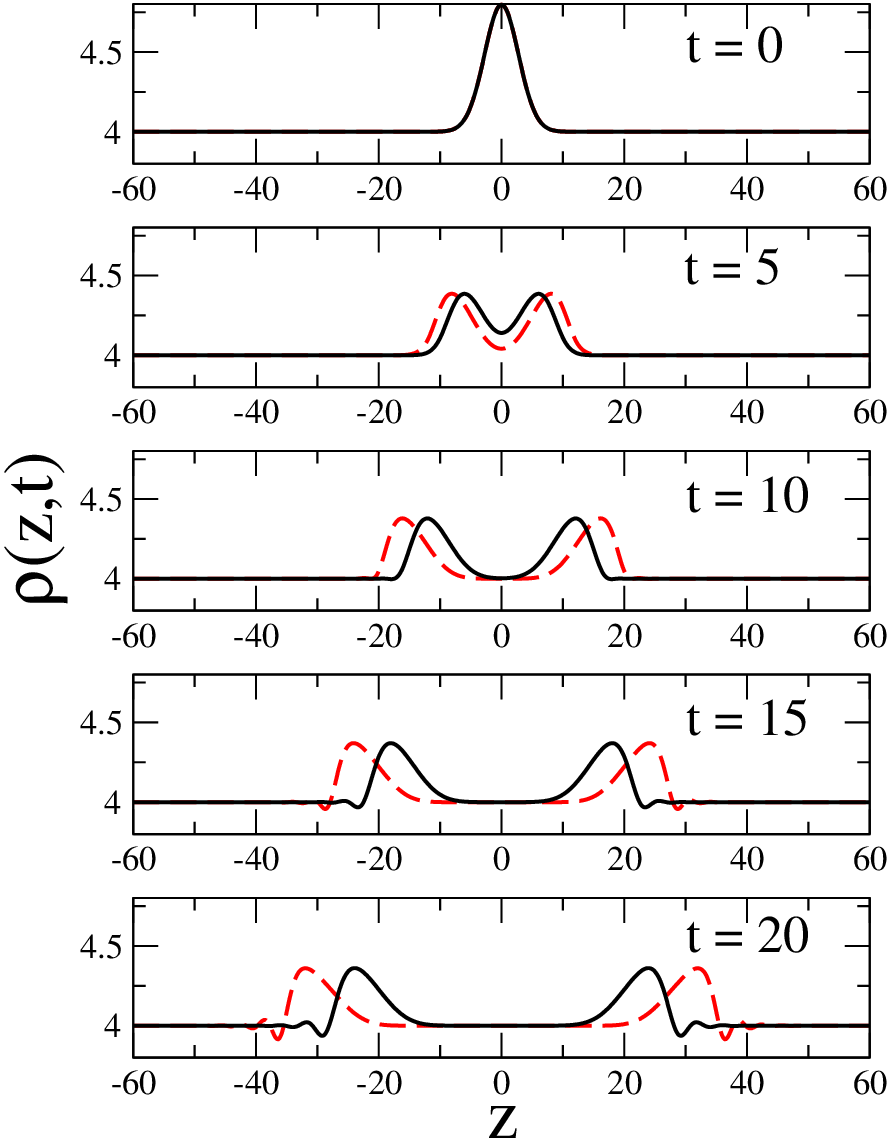}}
\caption{Time evolution of shock waves. 
Initial condition given by Eq. (\ref{initial-rho}) 
with $\sigma=4$, $\eta=0.2$ and $\rho_0=4$. 
The curves give the relative density profile $\rho(z,t)$ vs $z$ 
at subsequent times $t$. Solid lines obtained solving 
the time-dependent 1D NPSE, Eq. (\ref{tdnpse}), and dashed lines 
obtained solving the time-dependent 1D GPE, Eq. (\ref{tdgpe}). We choose 
the nonlinear strength $g=0.5$.} 
\label{fig3}
\end{center}
\end{figure}

In Fig. \ref{fig3} we plot the time evolution of shock waves 
obtained with $\rho_0=4$, $\sigma=4$, and $\eta=0.2$. The nonlinear 
strength is choosen as $g=0.5$, such that $g\rho=2$ and consequently 
the system is very far from the Tonks-Girardeau regime 
(characterized by $g\rho \ll g^2$). 
The figure displays the density profile $\rho(z,t)=|\psi(z,t)|^2$ 
at subsequent times. Note the splitting 
on the initial bright wave packet into two bright travelling waves moving 
in opposite directions. As previously discussed, 
there is a deformation of the two waves with the formation 
of a quasi horizontal shock-wave front. Eventually, 
this front spreads into dispersive wave ripples. 
The figure shows that there is no qualitative difference between 
1D NPSE (solid lines) and 1D GPE (dashed lines) 
in the physical manifestation of supersonic shock waves. 
Nevertheless, due to the different equation 
of state, there are quantitative differences. 
Our numerical simulation confirms that the velocity of the 
maximum of the shock wave is larger for the 1D GPE with respect 
to the 1D NPSE. 

\begin{figure}
\begin{center}
{\includegraphics[width=7.cm,clip]{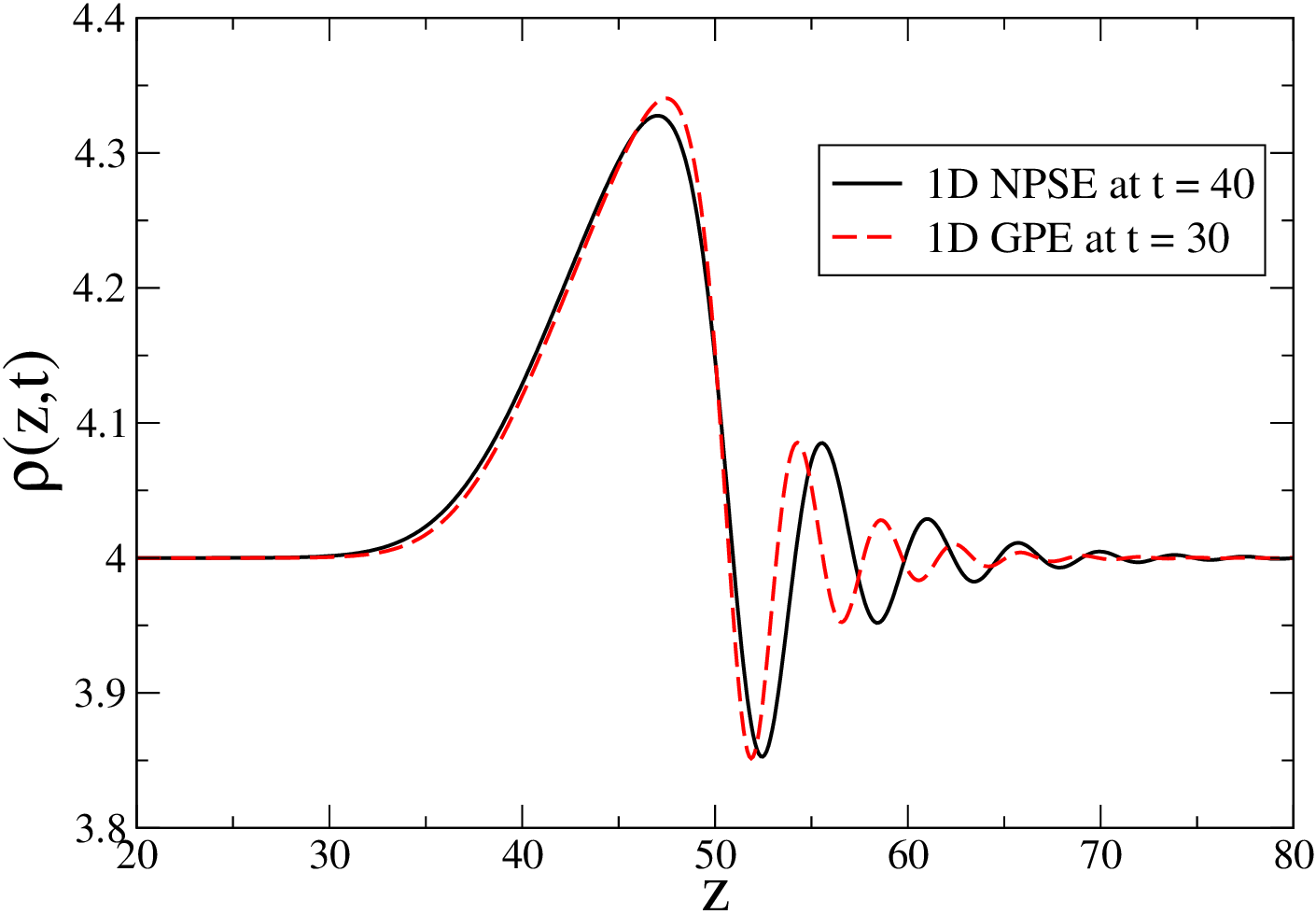}}
\caption{Zoom of the density profile $\rho(z,t)$ vs $z$ 
of the shock wave. Initial condition given by Eq. (\ref{initial-rho}) 
with $\sigma=4$, $\eta=0.2$ and $\rho_0=4$. Solid line obtained solving 
the time-dependent 1D NPSE, Eq. (\ref{tdnpse}) at $t=40$, and dashed line 
obtained solving the time-dependent 1D GPE, Eq. (\ref{tdgpe}) at $t=30$. 
We choose the nonlinear strength $g=0.5$.} 
\label{fig4}
\end{center}
\end{figure}

In Fig. \ref{fig4} we plot a zoom of the density profile $\rho(z,t)$ 
of the shock wave to better show the after-shock dispersive ripples. 
In the figure we compare the density profiles obtained 
with 1D NPSE and 3D GPE by chosing $t=30$ for 1D GPE and $t=40$ 
for 1D NPSE: in this way the spatial position of the highest maxima  
of the two profiles is practically superimposed. Fig. 4 clearly shows 
that the wavelength of the dispersive ripples is larger 
in the case of the 1D NPSE. We have verifed that this is a general feature 
by performing other numerical runs with different initial conditions. 

\section{Conclusions} 
 
We have shown that the shock-wave dynamics (velocity and density ripples) 
in a quasi 1D BEC, described by the 1D nonpolynomial Schr\"odinger 
equation (NPSE), is quite different with respect to the shock-wave dynamics 
of a strictly-1D BEC, described by the 1D Gross-Pitaevskii equation (GPE). 
The dynamics of the quasi 1D BEC becomes becomes equivalent to the one of 
the strictly-1D BEC only under the condition $g\rho \ll 1$,  
with $g$ the interaction strength and $\rho$ the one-dimensional 
axial density. In particular we have found that by using 1D NPSE 
the maximum of the shock wave moves slower, 
the time of shock formation is longer, and dispersive ripples 
have larger wavelength with respect to 1D GPE. For both 1D NPSE and 1D GPE 
we have obtained analytical and numerical solutions which 
could be compared with experimental data. Indeed, shock waves can be 
experimentally produced by engineering the initial density of the superfluid. 
Working with BECs made of alkali-metal atoms and 
confined in the transverse direction by a blue-detuned axial 
laser beam, a choosen axial density profile can be obtained by using a 
blue-detuned (bright perturbation) or a red-detuned (dark perturbation) 
laser beam perpendicular to the longitudinal axial direction. 
In this way one can experimentally test the equation of state 
of the quasi one-dimensional Bose-Einstein condensate analyzing the 
dynamical properties of the generated shock waves. 

\section*{Acknowledgments}

The author acknowledges Dr. Bogdan Damski for useful e-discussions during the 
earlier stage of this work and Italian Ministry of Education, 
University and Research (MIUR) for partial support 
(PRIN Project 2010LLKJBX "Collective Quantum Phenomena: 
from Strongly-Correlated Systems to Quantum Simulators"). 
The author thank the organizers of the Workshop 
Dispersive Hydrodynamics: The Mathematics of Dispersive Shock Waves 
and Applications, Banff Research Station, 2015.  



\begin{thebibliography}{99}

\bibitem{landau} L.D. Landau and E.M. Lifshitz, {\it Fluid Mechanics}, 
chapt. 10, par. 101 (Pergamon Press, London, 1987). 

\bibitem{bec-review} L.P. Pitaevskii and S. Stringari,  
{\it Bose-Einstein Condensation} (Oxford Univ. Press, Oxford, 2003).  

\bibitem{pitaevskii} S. Giorgini, L.P. Pitaevskii, and S. Stringari, 
Rev. Mod. Phys. {\bf 80}, 1215 (2008).   

\bibitem{luca12} L. Salasnich and F. Toigo, Phys. Rev. A 
{\bf 78}, 053626 (2008); 
L. Salasnich, Laser Phys. {\bf 19}, 642 (2009). 

\bibitem{hau} Z. Dutton, M. Budde, C. Slowe, and L.V. Hau, 
Science {\bf 293}, 663 (2001). 

\bibitem{cornell} M.A. Hoefer, M.J. Ablowitz, I. 
Coddington, E.A. Cornell, P. Engels, and V. Schweikhard, 
Phys. Rev. A {\bf 74}, 023623 (2006). 

\bibitem{hoefer} J.J. Chang, P. Engels, and M.A. Hoefer, 
Phys. Rev. Lett. {\bf 101}, 170404 (2008).

\bibitem{davis} R. Meppelink, S.B. Koller, J.M. Vogels, P. van der Straten, 
E.D. van Ooijen, N.R. Heckenberg, H. Rubinszein-Dunlop, S.A. Haine, 
and M.J. Davis, Phys. Rev. A {\bf 80}, 043606 (2009). 

\bibitem{thomas} J.A. Joseph, J.E. Thomas, M. Kulkarni, and A.G. 
Abanov A G, Phys. Rev. Lett. {\bf 106}, 150401 (2011). 

\bibitem{zak} I. Kulikov and M. Zak, Phys. Rev. A {\bf 67}, 063605 (2003). 

\bibitem{damski} B. Damski, Phys. Rev. A {\bf 69}, 043610 (2004); 
B. Damski, Phys. Rev. A {\bf 73}, 043601 (2006). 

\bibitem{gammal} A.M. Kamchatnov, A. Gammal, and 
R.A. Kraenkel, Phys. Rev. A {\bf 69}, 063605 (2004). 

\bibitem{perez} V.M. Perez-Garcia, V.V. Konotop, 
and V.A. Brazhnyi, Phys. Rev. Lett. {\bf 92}, 220403 (2004). 

\bibitem{muga} A. Ruschhaupt, A. del Campo, and J.G. Muga, 
Eur. Phys. J. D {\bf 40}, 399 (2006). 

\bibitem{noi} L. Salasnich, N. Manini, F. Bonelli, M. Korbman, 
and A. Parola, Phys. Rev. A {\bf 75}, 043616 (2007). 

\bibitem{noi2a} L. Salasnich, EPL {\bf 96}, 40007 (2011). 

\bibitem{noi2b} F. Ancilotto, L. Salasnich, and F. Toigo, 
Phys. Rev. A {\bf 85}, 063612 (2012); 
L. Salasnich, Few Body Sys. {\bf 54}, 697 (2013).  

\bibitem{bulgac} A. Bulgac, Y-L. Luo, and K.J. Roche, 
Phys. Rev. Lett. {\bf 108}, 150401 (2012).

\bibitem{npse} L. Salasnich, Laser Phys. {\bf 12}, 198 (2002);
L. Salasnich, A. Parola, and L. Reatto,
Phys. Rev. A {\bf 65}, 043614 (2002);
L. Salasnich, J. Phys. A: Math. Theor. {\bf 42}, 335205 (2009).

\bibitem{gll} L. Salasnich, A. Parola, and L. Reatto, 
Phys. Rev. A {\bf 70}, 013606 (2004);  
L. Salasnich, A. Parola, and L. Reatto, 
Phys. Rev. A {\bf 72}, 025602 (2005). 

\bibitem{ll} E.H. Lieb and W. Liniger, Phys. Rev. {\bf 130}, 1605 (1963). 

\bibitem{tg} M. Girardeau, J. Math. Phys. {\bf 1}, 516 (1960). 

\bibitem{spagnoli} A. Munoz Mateo and V. Delgado,
Phys. Rev. A {\bf 75}, 063610 (2007);
A. Munoz Mateo and V. Delgado, Phys. Rev. A {\bf 77}, 013617 (2008);
A. Munoz Mateo and V. Delgado, Ann. Phys. {\bf 324}, 709 (2009).

\bibitem{kamchatnov} A.M. Kamchatnov and V.S. Shchesnovich, 
Phys. Rev. A {\bf 70}, 023604 (2004). 

\bibitem{comesempre} L. Salasnich, Int. J. Mod. Phys. B 
{\bf 14}, 1 (2000).

\bibitem{toscani} P. Massignan and M. Modugno, 
Phys. Rev. A {\bf 67}, 023614 (2003); 
M. Modugno, C. Tozzo, and F. Dalfovo, 
Phys. Rev. A {\bf 70}, 043625 (2004); 
C. Tozzo, C.M. Kramer, and F. Dalfovo, 
Phys. Rev. A {\bf 72}, 023613 (2005); 
M. Modugno, Phys. Rev. A {\bf 73}, 013606 (2006). 

\bibitem{tedeschi} 
G. Theocharis, P.G. Kevrekidis, M.K. Oberthaler, and D.J. 
Frantzeskakis, Phys. Rev. A {\bf 76}, 045601 (2007); 
A. Weller, J.P. Ronzheimer, C. Gross, J. 
Esteve, M.K. Oberthaler, D.J. Frantzeskakis, G. Theocharis, 
and P.G. Kevrekidis, Phys. Rev. Lett. {\bf 101}, 130401 (2008). 

\bibitem{npse2} G. Mazzarella and L. Salasnich, 
Phys. Rev. A {\bf 82}, 033611 (2010). 

\bibitem{giovanni} 
G Mazzarella, M Moratti, L Salasnich, M Salerno, and F Toigo, 
J. Phys. B: At. Mol. Opt. Phys. {\bf 42}, 125301 (2009); 
G. Mazzarella, L. Salasnich, A. Parola, 
and F. Toigo, Phys. Rev. A {\bf 83}, 053607 (2011). 

\bibitem{luca-cs} L. Salasnich, A. Parola, and L. Reatto, 
Phys. Rev. A {\bf 69}, 045601 (2004). 

\bibitem{phase} S. Burger {\it et al.}, 
Phys. Rev. Lett. {\bf 83}, 5198 (1999); J. Denschlag 
{\it et al.}, Science {\bf 287}, 97 (2000). 

\bibitem{weiss} T. Kinoshita, T. Wenger,
and D.S. Weiss, Science {\bf 305}, 1125 (2004). 

\bibitem{sala-numerics} E. Cerboneschi, R. Mannella, E. Arimondo, 
and L. Salasnich, Phys. Lett. A {\bf 249}, 495 (1998); 
G. Mazzarella and L. Salasnich, Phys. Lett. A {\bf 373} 4434 (2009). 

\end{thebibliography}
\end{document}